\documentclass[12pt,a4paper]{article}

\usepackage[british]{babel}

\usepackage[a4paper,top=2cm,bottom=2cm,left=2.5cm,right=2.5cm,marginparwidth=1.75cm]{geometry}

\usepackage{cite}
\usepackage{amsmath}
\usepackage{graphicx}
\usepackage[colorlinks=true, allcolors=blue]{hyperref}
\usepackage{hyperref}
\usepackage[title]{appendix}
\usepackage{mathrsfs}
\usepackage{amsfonts}
\usepackage{booktabs} % For \toprule, \midrule, \botrule
\usepackage{caption}  % For \caption
\usepackage{threeparttable} % For table footnotes
\usepackage{algorithm}
\usepackage{algorithmicx}
\usepackage{algpseudocode}
\usepackage{listings}
\usepackage{enumitem}
\usepackage{chngcntr}
\usepackage{booktabs}
\usepackage{lipsum}
\usepackage{subcaption}
\usepackage{authblk}
\usepackage[T1]{fontenc}    % Font encoding
\usepackage{csquotes}       % Include csquotes
\usepackage{diagbox}
\usepackage{comment}
\usepackage{bbm}
\usepackage{float}

\usepackage{setspace}
\usepackage{titlesec}
\titleformat{\section} % Redefine section numbering format
  {\normalfont\Large\bfseries}{\thesection.}{1em}{}
  
\usepackage{lineno} % Add the lineno package

\rightlinenumbers % Right-align line numbers

\linenumbers % Enable line numbering

\usepackage{float}   % for customizing caption position
\usepackage{caption} % for customizing caption format


\title{Evaluation of respiratory disease hospitalisation forecasts using synthetic outbreak data}

\author[1,*]{Grégoire Béchade}
\author[2]{Torbjörn Lundh}
\author[3]{Philip Gerlee}
\affil[1]{\small Ecolé Polytechnique, \texttt{gregoire.bechade@polytechnique.edu}}
\affil[2]{\small Department of Mathematical Sciences, Chalmers University of Technology \& University of Gothenburg, Sweden, \texttt{torbjorn.lundh@chalmers.se}}
\affil[3]{\small Department of Mathematical Sciences, Chalmers University of Technology \& University of Gothenburg, Sweden \texttt{gerlee@chalmers.se}}
\affil[*]{\small Corresponding author: Grégoire Béchade}

\date{}  % Remove date

\begin{document}
\maketitle

\begin{abstract}
Forecasts of hospitalisations of infectious diseases play an important role for allocating healthcare resources during epidemics and pandemics. Large-scale analysis of model forecasts during the COVID-19 pandemic has shown that the model rank distribution with respect to accuracy is heterogeneous and that ensemble forecasts have the highest average accuracy. Building on that work we generated a maximally diverse synthetic dataset of 324 different hospitalisation time-series that correspond to different disease characteristics and public health responses. We evaluated forecasts from 14 component models and 6 different ensembles. Our results show that component model accuracy was heterogeneous and varied depending on the current rate of disease transmission. Going from 7 day to 14 day forecasts mechanistic models improved in relative accuracy compared to statistical models. A novel adaptive ensemble method outperforms all other ensembles, but is closely followed by a median ensemble. We also investigated the relationship between ensemble error and variability of component forecasts and show that the coefficient of variation is predictive of future error. Lastly, we validated the results on data from the COVID-19 pandemic in Sweden. Our findings have the potential to improve epidemic forecasting, in particular the ability to assign confidence to ensemble forecasts at the time of prediction based on component forecast variability.
\end{abstract}


%-------------------------------------------
% Paper Body
%-------------------------------------------
%--- Section ---%
\section*{Significance statement}
Accurate forecasts of hospitalisations during infectious disease outbreaks are essential for effective healthcare planning. However, during the COVID-19 pandemic, many forecasting models performed inconsistently, and it remains unclear which models work best under varying conditions. We developed a novel approach using synthetic data to systematically evaluate the accuracy of 14 forecasting models and six ensemble methods across a broad range of outbreak scenarios. Our results show that forecast performance depends strongly on the stage of the epidemic and that an adaptive ensemble method can reliably outperform others. This work improves our understanding of when and why specific forecasting methods succeed and provides practical tools to enhance epidemic preparedness and response in future outbreaks.
\section{Introduction}
%Forecasting in infectious disease epidemiology: forecast vs. scenario
Prediction of future outcomes such as incidence, hospital admissions and mortality, plays an important role in infectious disease epidemiology \cite{lutz2019applying}, both in terms of short-term prediction or forecasts and longer-term scenario projections. The time-scales relevant for forecasting and projections depends on the characteristics of the disease, but it is generally acknowledged that the accuracy of forecasts degrade rapidly beyond a month \cite{cramer2022evaluation}. This depends both on the inherently non-linear nature of disease transmission and on potential rapid changes in processes relevant for transmission, e.g. altered contact rates due to enforced or voluntary social distancing \cite{fox2022real,gerlee2021predicting}. 

%The utility of short-term forecasts for hospital planning: covid & flu
Forecasts of key epidemiological variables have played an important role in a number disease outbreaks including Zika virus \cite{kobres2019systematic}, Ebola \cite{munday2024forecasting}, Seasonal Influenza \cite{reich2019collaborative} and COVID-19 \cite{cramer2022evaluation}. The utility of infectious disease forecasts ranges from informing decision-makers about the future developments of the epidemic, to the distribution of medical supplies and supporting the allocation of health-care resources.

Forecasting models require historical data in order to produce useful forecasts. While mortality data typically remains stable during an epidemic it is often subject to reporting delays and in addition there is typically a delay of several weeks from infection to death, which can hamper forecasting efforts. Incidence data on the other hand does not suffer from this drawback, but is instead affected by testing strategies, which can vary significantly during the course of an epidemic. It has been argued that hospital admission present a useful middle-ground since admission criteria are more stable and forecasts of hospitalisations are valuable for regional decision-makers and allocation of healthcare resources, such as ICU-beds \cite{fox2022real}.

%Ensembles are useful
For these purposes a whole range of different forecasting models have been utilised, ranging from statistical models, mechanistic/compartmental models in terms of ordinary differential equations to agent-based models \cite{nixon2022evaluation}. This diversity of approaches became obvious during the COVID-19 pandemic and was particularly highlighted through the efforts of a number of forecast hubs where research groups and individuals alike could submit forecasts on a regular basis as well as documentation and code of their model \cite{reich2022collaborative}. Most prominent of these initiatives was the US COVID-19 Forecast Hub \cite{Cramer2021} which operated from spring 2020 until the autumn 2023 and focused on forecasts of mortality and hospitalisations at both the national and state level in the US.  

An evaluation carried out on forecasts of mortality submitted to the US Forecast Hub showed large heterogeneity in performance among the 27 models that were considered \cite{cramer2022evaluation}. Approximately two-thirds of the models performed better than a naïve baseline model and no single model outperformed the others. Instead it was observed that an ensemble of model forecasts performed best on average. The hub ensemble was to begin with formed by taking an unweighted average of point predictions and quantile levels of the submitted forecasts, but was later changed into an ensemble formed from the median of the component forecasts. 

Other methods for forming an ensemble have also been considered, e.g. optimising the weights of the models with respect to the performance of component models on historical data \cite{ray2021challenges}. While this improves performance relative to an unweighted mean the performance is similar to a median ensemble \cite{brooks2020comparing}, which is both conceptually easier to communicate and computationally easier to calculate. However, there is currently no theoretical basis for choosing component models in an optimal way, nor a theoretical understanding of why certain ensembles perform better than others.

%Some large scale studies, e.g. Forecast hub, but correlated data
Thus, there is a need to better understand the performance of both individual models and ensemble forecasts. One approach to this problem is to evaluate models and ensembles in a large-scale epidemic dataset, and while the evaluation of the US Forecast Hub contains large number of individual forecasts they are not independent since they concern geographical locations with similar transmission dynamics. Our goal here is to develop a novel method for evaluating epidemiological forecasting models, that utilises synthetically generated diverse data that reflects both different disease characteristics and responses to the epidemic in terms of time-dependent mobility.

In order to further our understanding of epidemic forecasting we present a large-scale and maximally diverse dataset of hospital admissions against which forecast models can be tested. We have characterised the performance of 14 different component models ranging from statistical to mechanistic models and covering both univariate and multivariate models. We also evaluated six different ensemble models and present a novel adaptive ensemble which utilises information about the current transmission dynamics. Lastly, our large-scale approach offers new insight into the relationship between the error of ensemble forecasts and the variance of component forecasts, information that can be used to judge the reliability of ensembles.

%--- Section ---%
\section{Results}
\subsection{Overview of the approach}
In order to generate synthetic data on hospitalisations from an epidemic we made use of CovaSim \cite{kerr2021covasim} an agent-based model of COVID-19 transmission, where disease and transmission parameters can be tuned to represent a wide-range of respiratory tract infections with e.g. different severity profiles, serial interval distributions and fraction of asymptomatics. The transmission dynamics can also be modulated by specifying time-dependent mobility rates.

In the simulation we consider all cases that are severe and critical to require hospital admission and extract the number of hospitalised patients each day of the simulation.  In addition to hospitalisations we also record the incidence, the effective reproduction number and mobility each day.

\begin{figure}[H]
\centering
\includegraphics[width=1\linewidth]{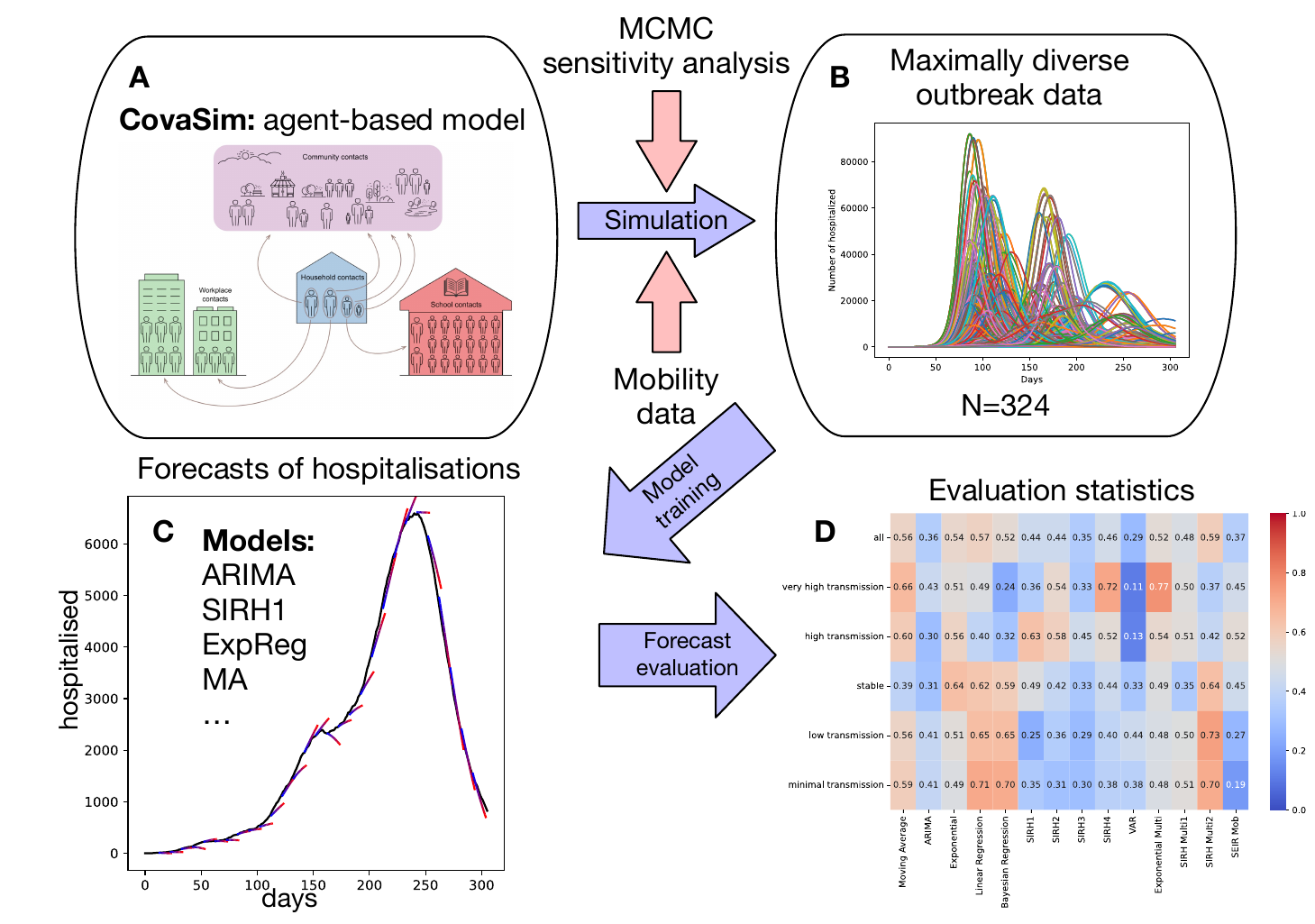}
\caption{\label{fig:overview}Overview of the methodology. A) We identified parameters in CovaSim, which when varied, gave rise to the most diverse set of outbreak data. B) Using these parameters we generated 324 different daily time-series of the number of hospitalised patients each X days long. C) 13 different forecasting models were trained on each time-series and the accuracy of 7 and 14 days ahead predictions was evaluated using RMSE and WIS. D) Results were aggregated and the models were ranked based on accuracy.}
\end{figure}

To generate a maximally diverse set of epidemics we constructed a metric that quantifies the difference between hospitalisation curves and made use of a Markov Chain Monte Carlo-method for identifying the subset of parameters, which when varied give rise to diverse epidemics (see figure \ref{fig:overview}A and Methods). We identified four parameters (symptomatic-to-severe rate, asymptomatic-to-recovered rate, probability of being symptomatic, probability of becoming severe) that when varied had the largest impact on the epidemic. We ran CovaSim with all possible parameters combinations when each parameter was either halved, at baseline or doubled. In addition we considered four different time-dependent mobility rates: seasonal, lockdowns, empirical and constant. This yielded 324 unique epidemics each 300 days long (see figure \ref{fig:overview}B). 

On each epidemic we trained 14 different forecast models and every 20 days we made forecasts of the number of hospitalised cases 7 and 14 days ahead (see figure \ref{fig:overview}C). We considered statistical models (e.g. exponential regression), autoregressive models (e.g. ARIMA) and mechanistic models (e.g. a SIRH-model which is an extension of the SIR-model with a compartment for hospitalised cases). Some models were univariate and only used the past hospitalisations to forecast the future, whereas others were multivariate and used past mobility and incidence for forecasting (e.g. a multivariate exponential regression and a VAR-model). We refer to the Methods for a detailed description of all models. The performance of the models was evaluted using both root mean squared error (RMSE) and the weighted interval score (WIS) \cite{bracher2021evaluating} (see Methods).

%\clearpage

\subsection{Performance of individual models}
\begin{figure}[H]
\centering
\includegraphics[width=1\linewidth]{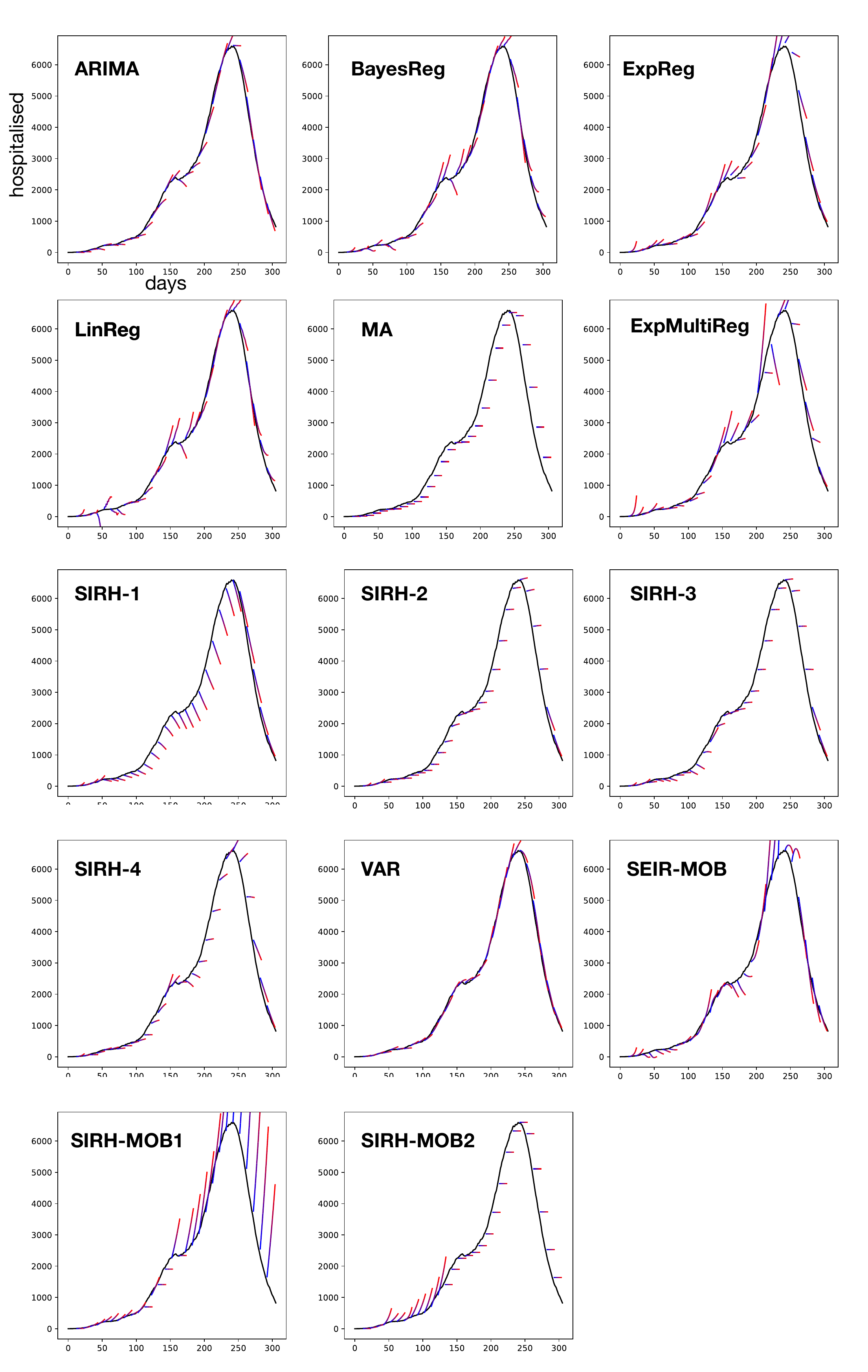}
\caption{\label{fig:forecasts}Forecasts of all models on a single example epidemic. Forecast are made every 10 days, starting at day 10, and stretch 14 days ahead. The colouring of the forecast curves indicates the length of the forecast (blue at day 0 and red at day 14).}
\end{figure}

Point forecasts of all considered models on an example epidemic are shown in figure \ref{fig:forecasts}. Here it can be seen that some models appear to provide accurate point forecast (e.g. the time-series ARIMA- and VAR-models), whereas others such as the SIRH-1 and ExpMultiReg fail to provide good forecasts.

\begin{figure}[!htb]
\centering
\includegraphics[width=0.8\linewidth]{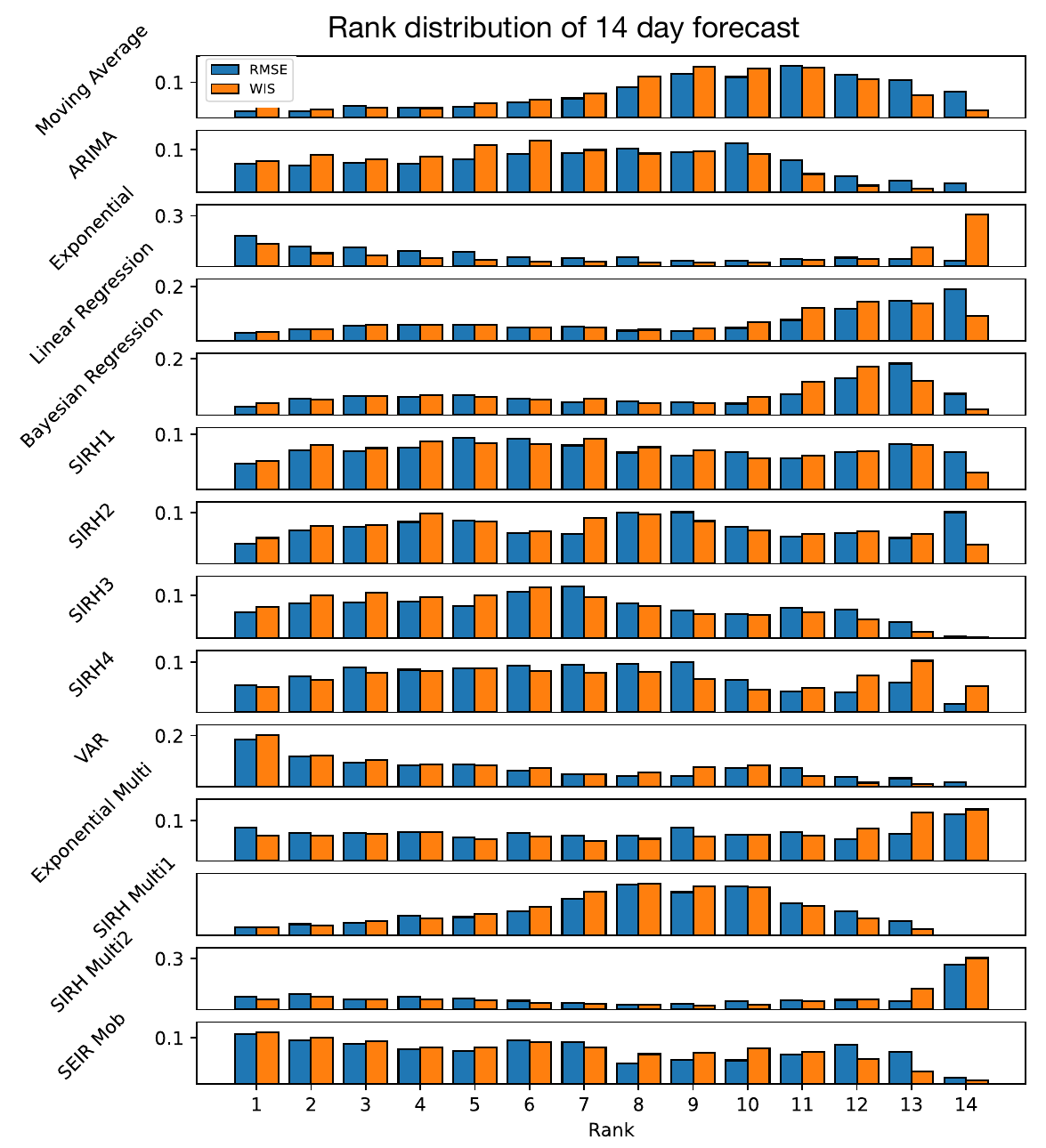}
\caption{\label{fig:ranks}Distribution of rankings of the models for all points for 14-day forecasts with respect to both RMSE and WIS. The average ranks for each model is reported in Table \ref{tab:ranks}}
\end{figure}

In order to systematically investigate the accuracy of the models across all 324 epidemics we calculated the RMSE and WIS of each forecast and ranked the models according to their performance. The distribution of ranks based on RMSE and WIS is shown in figure \ref{fig:ranks}, where height of the bars correspond to the fraction of times each model achieved the corresponding rank. The Moving Average (MA) model represents our baseline model and forecasts the average number of hospitalised cases in 7 day period prior to the day of forecast. As expected this model performs poorly with respect to both the WIS and RMSE metrics, although it does perform best in a small fraction of cases. In fact for most models the rank distribution with respect to WIS and RMSE are similar, the exception being the Exponential Regression model. This discrepancy arises because the ExpReg-model provides accurate point predictions, but (at times) excessively wide confidence intervals, which increases the WIS and places the model at the bottom of the ranking. The rank distribution for 7 day forecast is similar (see Supplementary fig. 1).

The rank distribution can be summarised by considering the average ranks of the models, which is shown in Table \ref{tab:ranks}. Here we have also normalised the ranks so that 0 corresponds to the best possible rank and 1 to the worst. To begin with note that the Moving Average (MA) model has the worst average normalised rank in all cases except for WIS at 14 days where Linear Regression and SIRH-Multi2 perform worse. We can see that the top performing models with respect to RMSE for 7 days is VAR, ExpReg and ARIMA, and for 14 days we have ExpReg, VAR and SEIR-Mob. If we instead consider WIS the top performing models for 7 days are VAR, ARIMA and SIRH3, and for 14 days forecasts: VAR, SIRH3 and ARIMA. It is worth noting that the top models contain both statistical and mechanistic models. If we consider the change in normalised rank going from 7 to 14 days we see that all models that have an autoregressive structure (ARIMA, LinReg, BayesReg and VAR) perform worse for longer forecast horizons. This is true for both WIS and RMSE. In contrast, all mechanistic models perform better at 14 days compared to 7 days (for both RMSE and WIS). The latter also holds true for the Moving Average-model and the two exponential regression models. 

\begin{table}[!htb]
    \centering
\begin{tabular}{lrrrr}
\toprule
Model & RMSE (7 days) & WIS (7 days) & RMSE (14 days) & WIS (14 days) \\
\midrule
Moving Average & 0.69 & 0.62 & 0.61 & 0.56 \\
ARIMA & 0.37 & 0.31 & 0.43 & 0.36 \\
Exponential & 0.34 & 0.55 & 0.33 & 0.54 \\
Linear Regression & 0.50 & 0.47 & 0.60 & 0.57 \\
Bayesian Regression & 0.44 & 0.42 & 0.55 & 0.52 \\
SIRH1 & 0.50 & 0.47 & 0.47 & 0.44 \\
SIRH2 & 0.53 & 0.47 & 0.48 & 0.44 \\
SIRH3 & 0.43 & 0.36 & 0.40 & 0.35 \\
SIRH4 & 0.45 & 0.51 & 0.41 & 0.46 \\
VAR & 0.29 & 0.26 & 0.33 & 0.29 \\
Exponential Multi & 0.49 & 0.54 & 0.48 & 0.52 \\
SIRH Multi1 & 0.55 & 0.53 & 0.50 & 0.48 \\
SIRH Multi2 & 0.53 & 0.61 & 0.52 & 0.59 \\
SEIR Mob & 0.41 & 0.38 & 0.40 & 0.37 \\
\bottomrule
\end{tabular}
    \caption{Average normalised rank of all 14 models for 7- and 14-day forecasts with respect to RMSE and WIS.}
    \label{tab:ranks}
\end{table}

\begin{figure}[!htb]
\centering
\includegraphics[width=0.8\linewidth]{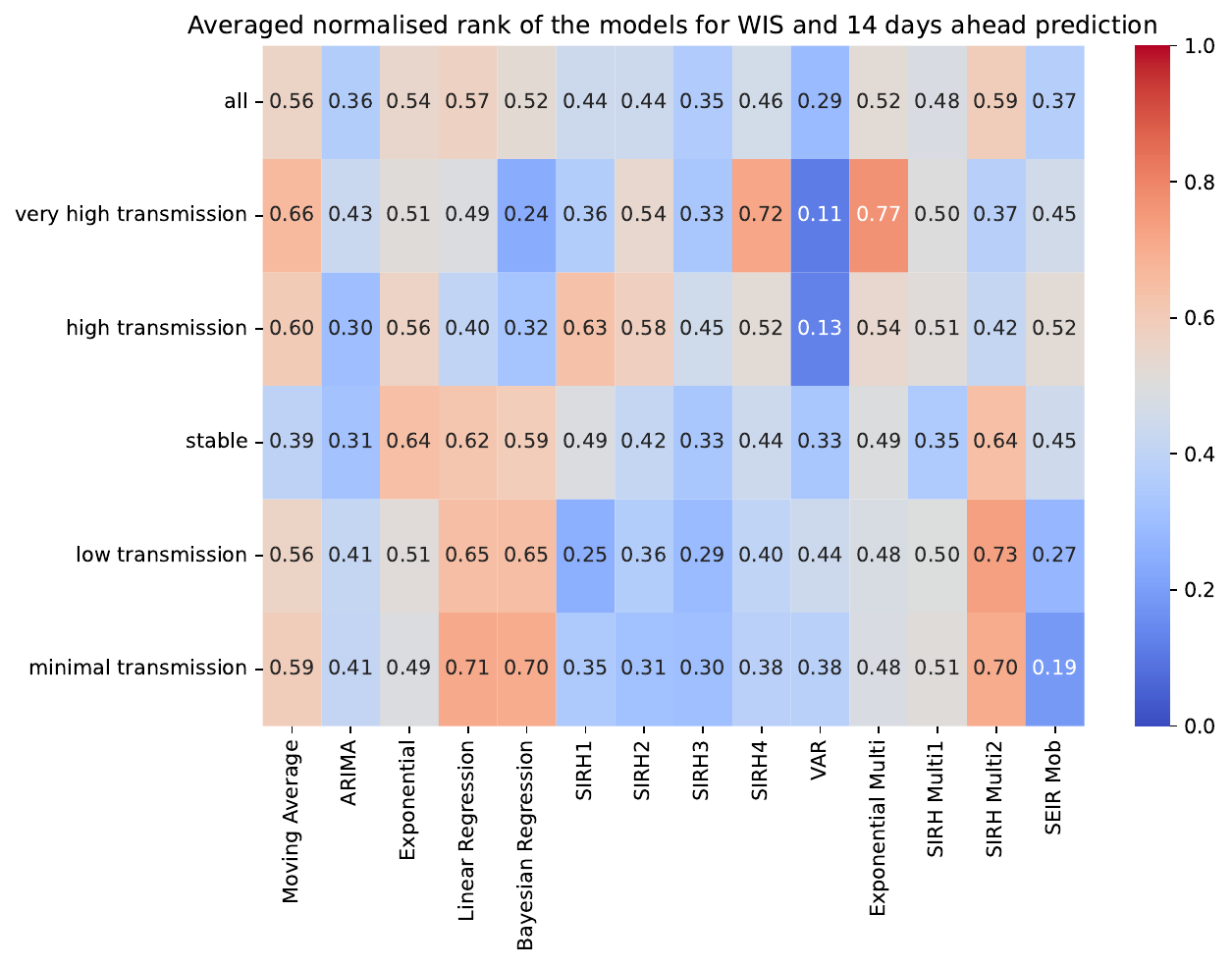}
\caption{\label{fig:model_perf}Heatmap of average normalised model rank based on WIS for 14-day forecasts. The day of forecast is classified according to the current effective reproduction number according to: minimal transmission ($R_{\textrm{eff}} < 0.5$), low transmission ($0.5\leq R_{\textrm{eff}} < 0.8$), stable ($0.8\leq R_{\textrm{eff}} < 1.2$), high transmission ($1.2\leq R_{\textrm{eff}} < 3$) and very high transmission $R_{\textrm{eff}}\geq 3$.}
\end{figure}

We then went on to stratify the average normalised ranks based on the characteristic of the epidemic in terms of the effective reproductive number $R_{\textrm{eff}}$ at the day the forecast was made. We consider five different classes ranging from minimal transmission ($R_{\textrm{eff}} < 0.5$), low transmission ($0.5\leq R_{\textrm{eff}} < 0.8$) to stable ($0.8\leq R_{\textrm{eff}} < 1.2$), high transmission ($1.2\leq R_{\textrm{eff}} < 3$) and very high transmission $R_{\textrm{eff}}\geq 3$. The results can be seen in figure \ref{fig:model_perf} which shows the average normalised rank using WIS at 14 days. It is clear that model performance in general is quite heterogeneous with respect to $R_{\textrm{eff}}$. For example the VAR-model performs very well at times of high and very high transmission (average rank of 2.6 and 2.8 respectively), but considerably worse at low transmission (average rank 6.3), where it is outperformed by six other models. On the other hand the SEIR-Mob model is the best model during minimal transmission (average rank 3.6), but is outperformed by seven other models during high transmission (results for WIS 7 days and RMSE for 7 and 14 day forecasts are similar, see Supplementary fig. 2). This suggest that knowledge about the current status of the epidemic can inform model choice, a fact we will return to when constructing ensemble models.  

\begin{figure}[!htb]
\centering
\includegraphics[width=0.8\linewidth]{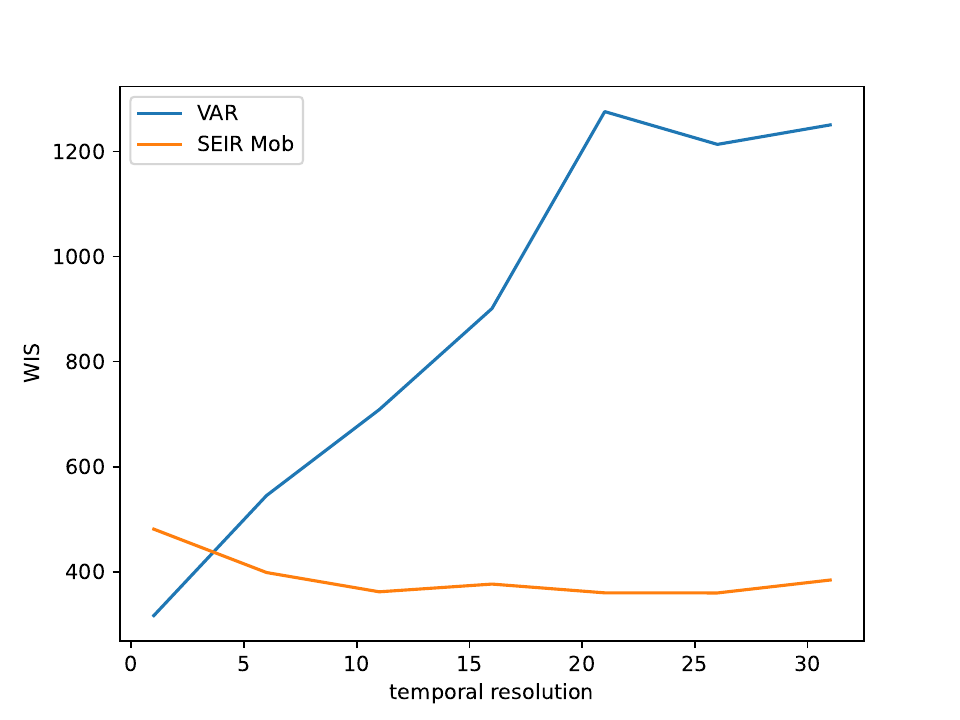}
\caption{\label{fig:resol}Performance of the VAR- and SEIR Mob-model as the temporal granularity is decreased. At temporal granularity of $T=1$ the models utilise daily prevalence and mobility data, whereas for higher granularity the data is available every $T$ days. For reference the average WIS of the baseline MA-model equals 1475.}
\end{figure}

As noted earlier some models make use of multivariate data to make forecasts of future hospitalisations. These additional data streams in terms of incidence and mobility are provided to the models at the same daily resolution as the hospitalisation data. In a sense this makes for an unfair comparison between univariate and multivariate models since incidence and mobility data rarely is available at such high temporal resolution. In order to investigate the impact of the temporal resolution of the additional data streams we reduced the resolution in a range from daily to every 30 days. Historical data was linearly interpolated and care was taken not to use data beyond the date of forecast during interpolation. We then calculated the performance in terms of average WIS across all epidemics. This was carried out for the VAR and SEIR-Mob models, which both performed well compared to the other models, and the result can be seen in figure \ref{fig:resol}. For the VAR-model, which uses both incidence and mobility, the performance decreases (corresponding to an increase in WIS) approximately linearly from a daily temporal resolution to a resolution of roughly 20 days at which point the performance saturates. For reference the average WIS of the Moving Average model equals 1475, which is above the WIS of the VAR-model for all resolutions. In contrast the SEIR-Mob model, which uses mobility data, performs even better (i.e. lower WIS) when the resolution is increased to 10 days and from then on the WIS remains roughly constant. A possible explanation for this is that for one of the mobility datasets the mobility changes on a weekly basis and the linear interpolation employed to reduce the resolution in fact improves the utility of the data. In conclusion, these results suggest that in realistic scenarios where incidence and mobility are available on a weekly resolution (or less) the mechanistic SEIR-Mob model is preferable compared to the autoregressive VAR-model.

\subsection{Ensemble model}
\begin{figure}[!htb]
\centering
\includegraphics[width=0.8\linewidth]{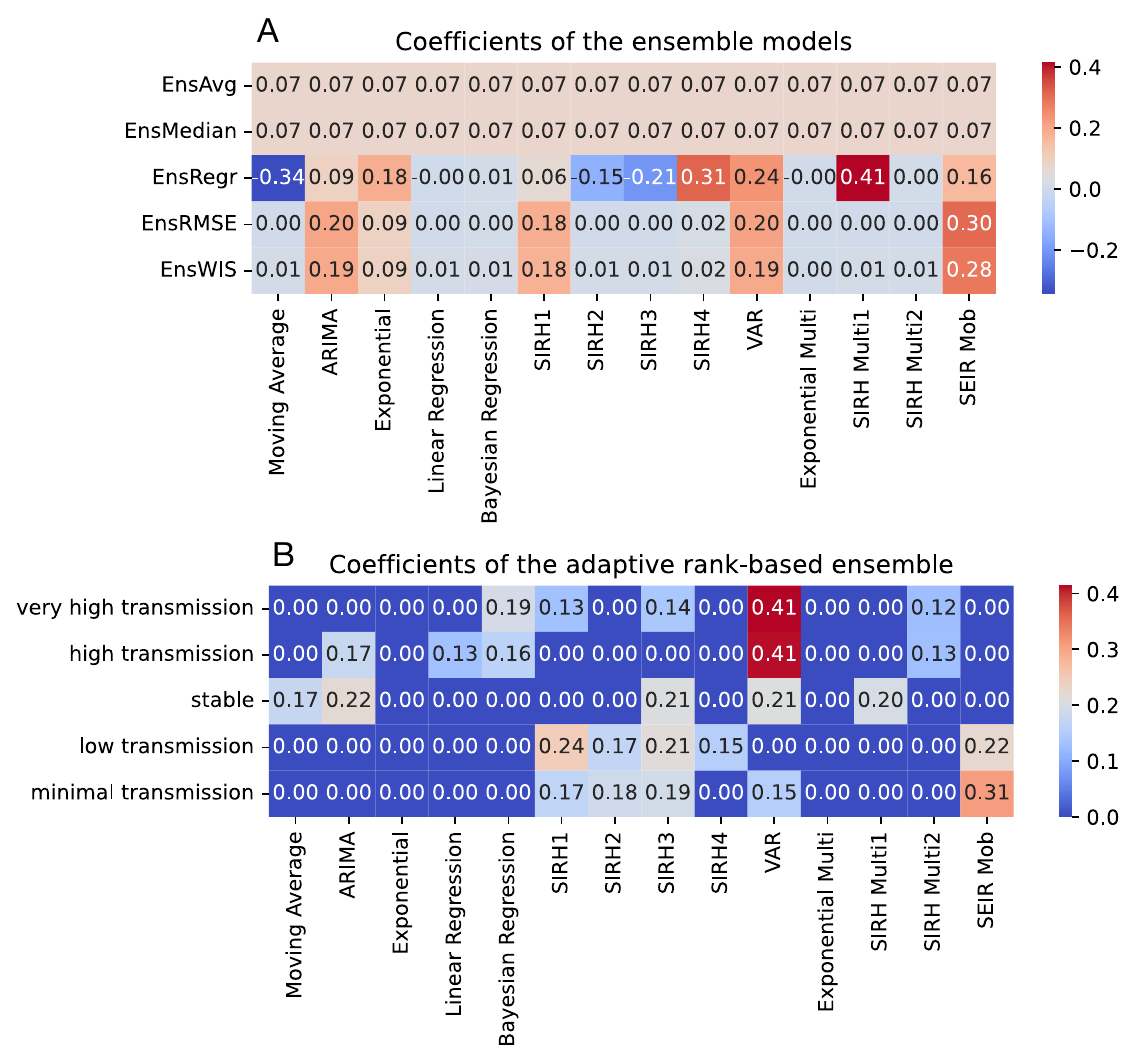}
\caption{\label{fig:ens_coeff} A. The weights of component model predictions in the ensembles. Note that for the Median Ensemble (EnsMedian) the point prediction is given as the median of the model predictions. This also applies to the quantiles of the EnsMedian-prediction. B. The weights in the rank-based adaptive ensemble that are choosen depending on the current effective reproductive number.}
\end{figure}

It has been established that ensemble forecasts typically outperform component models in terms of performance \cite{taylor2023combining}. We investigate this observation in the context of our large-scale synthetic dataset and consider the following types of ensembles:
\begin{itemize}
    \item Average Ensemble (EnsAvg): this ensemble takes an unweighted mean of all component models to calculate the point prediction and confidence intervals \cite{cramer2022evaluation}.
    \item Median Ensemble (EnsMedian):  this ensemble uses the median of all point predictions and confidence intervals \cite{ray2023comparing}.
    \item Regression Ensemble (EnsReg): this ensemble takes a weighted mean of the component point predictions and confidence intervals plus an intercept. The weights and intercept were obtained using linear regression where the component model forecasts are covariates and the actual hospitalisation is the target variable.
    \item RMSE-optimised Ensemble (EnsRMSE): same as EnsReg with the difference that the weights are constrained to sum to unity and there is no intercept \cite{brooks2020comparing}.
    \item WIS-optimised Ensemble (EnsWIS): same as EnsRMSE but here the weights are selected to minimise the WIS of the ensemble forecast. Again the weights sum to unity \cite{brooks2020comparing}.
    \item Rank-based Ensemble (EnsRank): this ensemble changes its weights depending on the current effective reproductive number. For each of the five regimes defined in fig.\ \ref{fig:model_perf} we pick the five models with the smallest average rank. The weights of the models are set to the inverse of the model rank, and lastly the weights are normalised to sum to unity, similar to what is done in \cite{taylor2023combining}.
\end{itemize}
The model weights were obtained by splitting the epidemics into a training and evaluation set and calibrating the weights on the training set. The resulting model weights are shown in fig. \ref{fig:ens_coeff}.

\begin{figure}[!htb]
\centering
\includegraphics[width=0.8\linewidth]{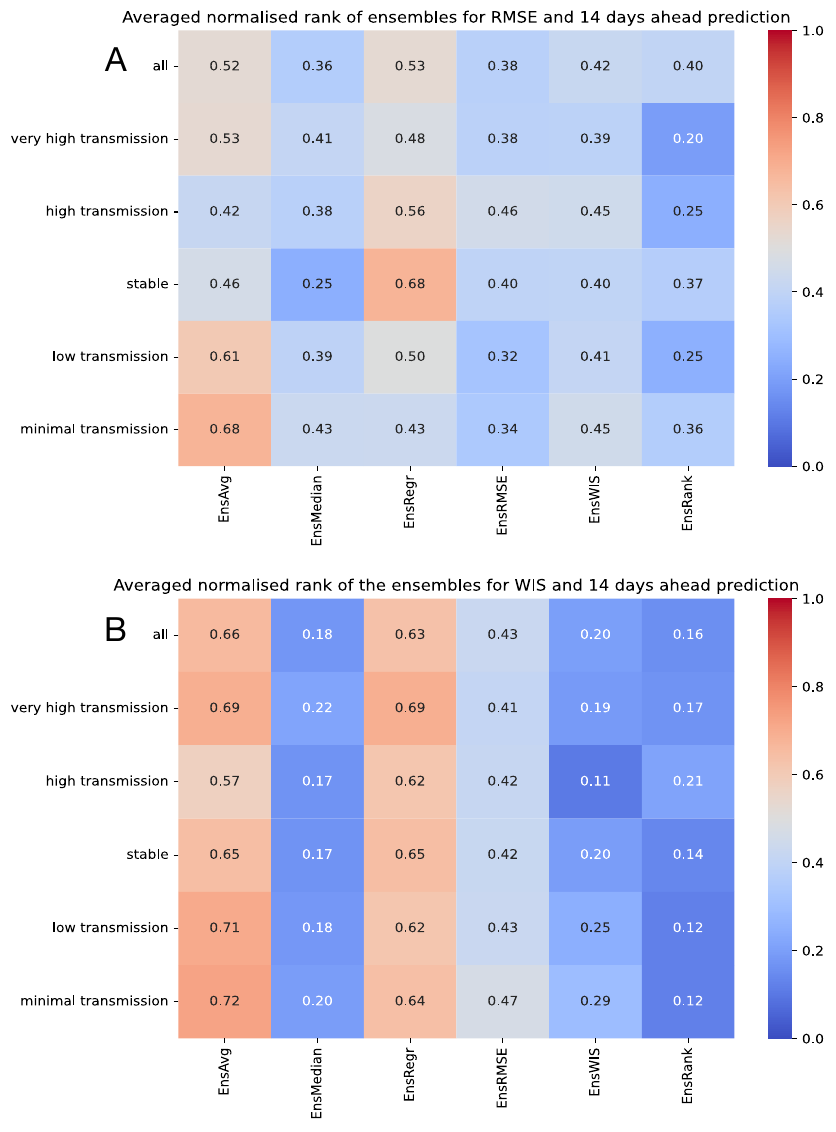}
\caption{\label{fig:ens_perf}The average normalised ensemble rank as compared to both component models and the six ensembles for A) RMSE and B) WIS.}
\end{figure}

We then evaluated the performance of the ensembles on the evaluation set with respect to both WIS and RMSE. The average normalised rank of the ensembles when compared to all models (in total 20 models, 14 component models plus 6 ensembles) is shown in fig. \ref{fig:ens_perf}. In terms of RMSE-performance the ensembles do not outperform the component models when considering all points, but do so for some types of points, e.g. EnsRank has the lowest average normalised rank for 'low transmission' and EnsMedian has the lowest average normalised rank when transmission is 'stable' (see Supplementary fig. 3 for a comparison of all models and ensembles). However, for WIS-performance the picture is different. Here three ensembles, EnsMedian, EnsWIS and EnsRank, clearly outperform all component models and other ensembles. EnsRank has the lowest overall rank for all transmission regimes except 'high transmission', where EnsMedian has the lowest average rank.

\begin{figure}[!htb]
\centering
\includegraphics[width=0.7\linewidth]{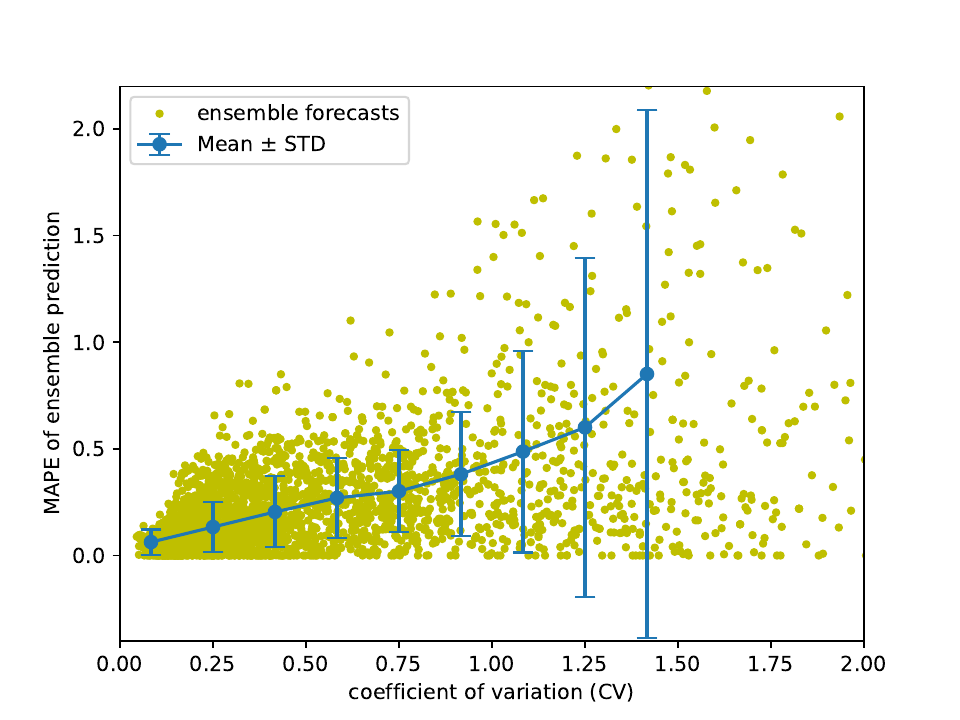}
\caption{\label{fig:cv}The coefficient of variation (CV) of the individual model predictions for 7-day forecasts and the corresponding error (MAPE) of the Median Ensemble prediction. The solid line shows the mean MAPE in each bin and the error bars correspond to one standard deviation (see Supplementary fig. 4 for the corresponding plot for 14 day forecasts).}
\end{figure}

In situations where component models in an ensemble agree it is reasonable to think that the ensemble should be more accurate compared to a situation where component predictions diverge. To investigate this hypothesis we measured the accuracy of the median ensemble in terms of the mean absolute percentage error (MAPE) and compared it to the coefficient of variation (CV) of the component forecast. CV is calculated as the sample standard deviation divided by the sample mean and is a normalised measure of variation within a sample. Figure \ref{fig:cv} shows a scatter plot of the MAPE and CV of each individual forecast for all epidemics together with the mean and standard deviation of  binned data. From the plot it is clear that a small CV implies a small MAPE, whereas for large CVs the MAPE can take a range of values. Although the data has a strong heteroscedastic trend we observe an increased mean MAPE as the CV of the ensemble increases. This implies that the CV of the ensemble is predictive of future error of the ensemble forecast. 

\subsection{Evaluation on real data}
\begin{figure}[!htb]
\centering
\includegraphics[width=1\linewidth]{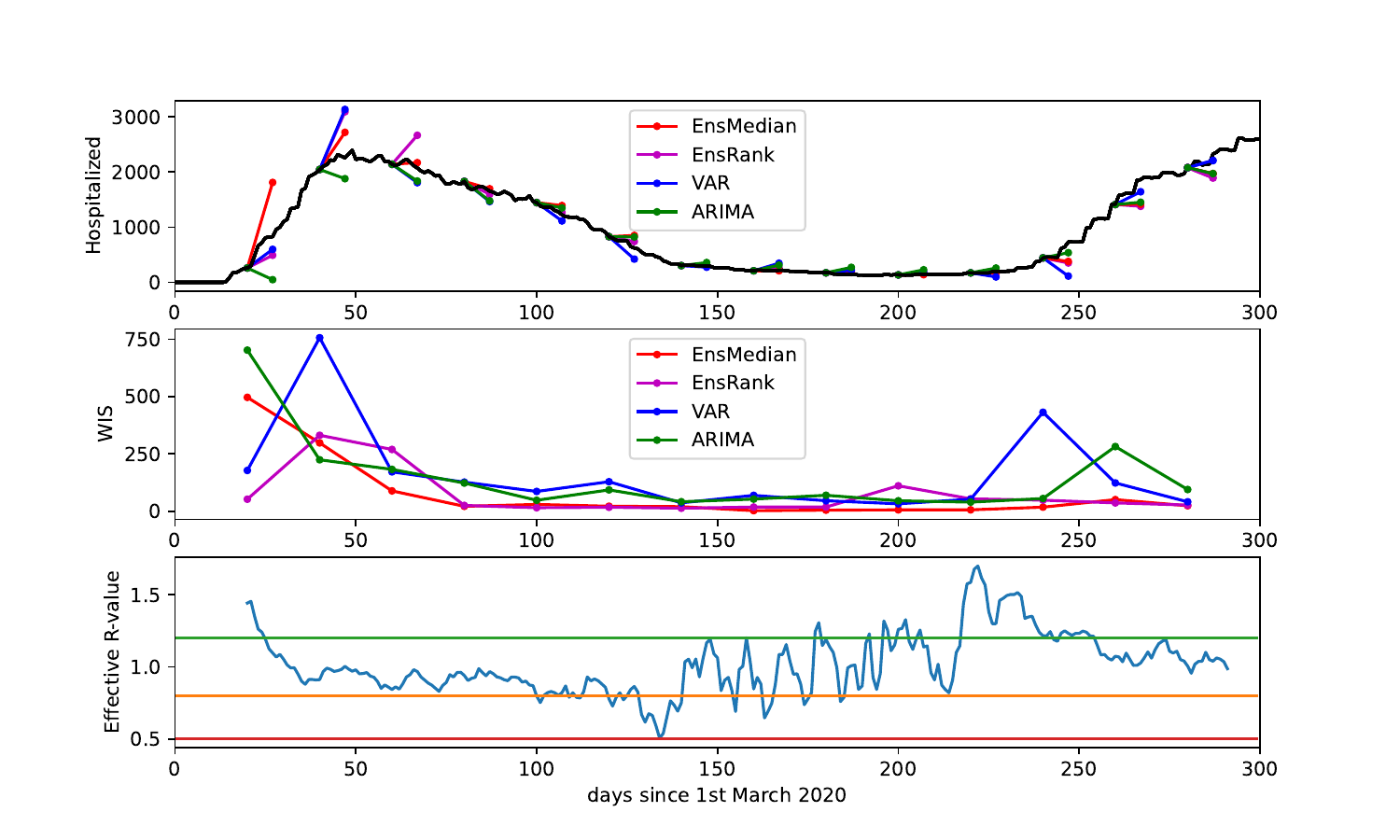}
\caption{\label{fig:sweden}Evaluation of selected component model and ensemble 7-day forecasts for COVID-19 data from Sweden during 2020. A) Point predictions from the best performing ensembles (median- and adaptive rank-based) and component models (VAR and ARIMA). B) The weighted interval score (WIS) of the ensembles and models calculated from prediction intervals for each forecast. C) Estimate of effective reproduction value. The solid lines correspond to break points for the adaptive rank-based ensemble. The red line corresponds to $R_{\textrm{eff}}=0.5$, the orange line to $R_{\textrm{eff}}=0.8$ and the green line to $R_{\textrm{eff}}=1.2$.}
\end{figure}

To validate our findings concerning the performance different ensembles we trained the models and made forecasts of hospitalised COVID-19 patients during 2020 in Sweden. We calculated the WIS for the top two component models (ARIMA and VAR) and ensembles (EnsMedian and EnsRank). The results are shown in fig. \ref{fig:sweden} together with the instantaneous reproductive number estimated from incidence data \cite{cori2013new},  which is used by the rank-based ensemble. Also here we see that the ensemble forecasts achieve lower WIS and EnsRank achieves a slightly lower WIS compared to EnsMedian (73.8 vs. 77.4).

\section{Discussion}
We have investigated the ability of a wide array of forecasting models to predict the number of hospitalised cases of a hypothetical respiratory infectious disease. A diverse set of outbreak data was generated using CovaSim, an established agent-based model of COVID-19 transmission, whose parameters were adjusted to generate a maximally diverse set of hospitalisation curves (see \ref{fig:overview}B). 

We considered 14 forecasting models that were either autoregressive, statistical or mechanistic and were univariate or multivariate (made use of multiple data types for prediction). The accuracy of the models was evaluated using both RMSE (for point predictions) or WIS (for probabilistic predictions) for 7- and 14-day forecasts. For each forecast we ranked the models according to their RMSE/WIS with the first ranked model having the lowest error. The resulting rank distributions (fig.\ \ref{fig:ranks}) based on all epidemics and forecasts (in total 4860 points) are highly heterogeneous were no model consistently outperforms the others. Similar findings have been made by forecast hubs during the COVID-19 pandemic \cite{cramer2022evaluation}.

In fact all models are at some point both the worst and best model. However, there are individual differences where some models have distributions shifted towards higher rank (e.g. Bayesian regression), whereas others are shifted towards low rank (e.g. the VAR-model). The rank distributions with respect to RMSE and WIS are in general similar, but Exponential regression is an exception with a distribution which is shifted towards lower rank with RMSE and higher rank with WIS. The reason behind this is that WIS punishes forecast with excessive confidence intervals, which the Exponential regression model at times produces.

Taking the average of the rank distributions and normalising (see Table \ref{tab:ranks}) we notice that almost all models outperform the Moving Average-model, which serves as a baseline model, the exception being Linear Regression and the SIRH-Multi2, which perform worse with respect to WIS at 14 day forecasts. The model performance relative to the baseline model found in this study is better than the one reported in \cite{cramer2022evaluation}, which possibly is related to the larger and more diverse dataset used for evaluation. 

Going from 7 to 14 day forecasts we note that all autoregressive models decrease their relative accuracy with respect to both RMSE and WIS, whereas the mechanistic models improve their accuracy. The two exponential regression models also show a slight improvement for longer forecasts. The relative improvement of mechanistic models is in line with previous studies that have shown that capturing transmission dynamics in the model structure improves long-term accuracy \cite{rahmandad2022enhancing}.

%\textbf{Has this been observed before?}

The heterogeneity of performance becomes even more evident when we stratified the rank based on the effective reproduction value at the time of forecast (see fig. \ref{fig:model_perf}). From this it can be seen that the autoregressive models in general perform better during high/very high transmission compared to low/minimal.

%temporal resolution, decay of VAR-model, linear interpolation helps SEIR.

This information could guide the use of different forecast models during an ongoing pandemic. For example it would be possible to evaluate the performance of any set of forecast models on synthetic data and stratify the accuracy based on effective reproductive number. The results could then inform decision-makers when considering multiple forecasts. 

As we have shown the stratification can also be used to form an adaptive rank-based ensemble, where the weights of component models are adjusted depending on the instantaneous reproduction number. This rank-based ensemble outperforms all other ensemble methods (see fig. \ref{fig:ens_perf}), although the median ensemble and WIS-optimised ensemble show similar performance. However, the good performance of the median ensemble, which has been reported previously \cite{brooks2020comparing}, together with its simplicity makes it a strong candidate for producing forecasts. In particular since it requires no adjustment of model weights based on historical data.

The application of the two best ensembles on real data from Sweden during the COVID-19 pandemic showed that the results from the synthetic data also hold on real hospitalisation data. We see that the median and rank-based ensemble both outperform the top two component models (with the lowest average rank for 7 day forecasts with WIS) and that the rank-based ensemble has a lower average WIS than the median ensemble. 

A benefit of our approach with a large-scale synthetic, yet realistic, data set is that the results we obtain are more reliable compared to those obtained from a single epidemic. In addition, it becomes possible to see trends in the results that are not realised in smaller datasets. One notable example of this is the relation between the coefficient of variation (CV) of component forecasts within an ensemble and the future error of the ensemble forecast (see fig. \ref{fig:cv}). That ensemble CV is indicative of forecast error has previously been observed in meteorology, where ensembles are used to account for both uncertainty in initial condition and model structure \cite{whitaker1998relationship}. In meteorology, this is known as the spread-skill relationship and has been shown to be informative about the expected quality of the forecast. By assuming a simple statistical model for the relationship between spread and error an upper bound on the correlation coefficient between the variables can be obtained. In our data we find that for CV in the range $0 < \mathrm{CV} < 2$ the correlation coefficient between CV and MAPE-error equals 0.51, whereas the theoretical upper bound, which assumes an unbiased ensemble, equals approximately 0.6 \cite{houtekamer1992quality}. However, it should be noted that the relationship between CV and error is highly heteroscedastic and further analysis and validation is required before the CV can be used operationally to assign confidence to ensemble forecasts. A similar relationship was observed by \cite{shaman2012forecasting} who made use of an ensemble adjustment Kalman filter to model influenza epidemics. They used a single model for disease transmission, and the ensemble was formed by multiple instances of the model with different states and parameter values. In that context they showed that that the logarithm of the ensemble variance was predictive of the ability of the ensemble to forecast the timing of the peak of the outbreak. This is similar to our results, but yet distinct since we consider a multi-model ensemble with 14 different component models.

%Limitations
This study has several limitations. To begin with we consider synthetic data generated using an agent-based model of respiratory disease transmission. Although this model accounts for many of the processes involved in disease transmission it might still miss out on important characteristics of real data. In addition we assumed perfect access to hospital data without any delays or misreporting. In realistic settings there is typically a reporting delay which reduces the effective forecasting horizon. Reporting delays can be handled using nowcasting methods \cite{bergstrom2022bayesian}, but this introduces additional uncertainty in recent data.

Secondly, we only consider a limited set of forecast models. Some of these models have been used during the COVID-19 pandemic to predict hospital admission \cite{kitaoka2023improved,gerlee2021predicting,alabdulrazzaq2021accuracy,kuchenhoff2021analysis}, but our study does not include other commonly used models such as age-structured compartmental models, machine learning models and agent-based models. Some of the models we consider make use of additional data streams in terms of incidence and mobility, but there are other data sources that can be informative when forecasting hospitalisations, e.g. telenursing calls \cite{spreco2022nowcasting} and waste-water data \cite{gudde2025predicting}.

Our study focused on hospitalised cases as the target variable, but there are a number of other variables that are of interest to decision-makers during an epidemic. For example, forecasting incidence can be useful since changes in incidence typically precede changes in hospitalisation, thus providing an early warning signal. Also, forecasting mortality has been considered useful since it provides an estimate of coming severity of an outbreak.  

The conclusions drawn from the results concerning different ensemble methods and the spread-skill relationship are based on the component models considered in this study. In order to strengthen those conclusion one would like to carry out similar studies with respect to both different component models and target data sets. Optimally this would be performed in a prospective setting on real data. Also, our results are limited to short-term forecasting and it would be interesting to see if similar results concerning ensembles are obtained for longer term forecasts or even scenario projections.

%Future work
In conclusion, this study, which is based on synthetic outbreak data, shows that forecast model performance is heterogeneous with respect to different outbreaks and disease transmission regime. We have shown that this information can be harnessed to form an adaptive rank-based ensemble, which outperforms traditional ensemble types. In future work we plan to investigate how well the rank-based ensemble generalises to other context and how the performance depends on the amount of training data. The spread.-skill relationship of ensembles that we have uncovered also requires more investigation, in particular in the context of other component models and datasets. 

\section{Methods}
In this section we provide details of the data generation, the forecast models, performance metrics and ensemble forecasts. All code and data are available at: \href{https://github.com/philipgerlee/Evaluation-of-respiratory-disease-hospitalisation-forecasts-using-synthetic-outbreak-data}{GitHub}

\subsection{Data generation -- CovaSim}
To generate the synthetic outbreak data CovaSim \cite{kerr2021covasim}, an agent-based model was used that can simulate the transmission of a respiratory infectious disease in a population. 
This model takes as an input disease- and population-specific parameters such as the population size, the age distribution, the transmission probability and age-dependent severity, 
and outputs a complete description of the outbreak, such as time-series of the number of severe and asymptomatic cases, and also  the effective reproduction number. 

\begin{figure}[!htb]
\centering
\includegraphics[width=0.7\linewidth]{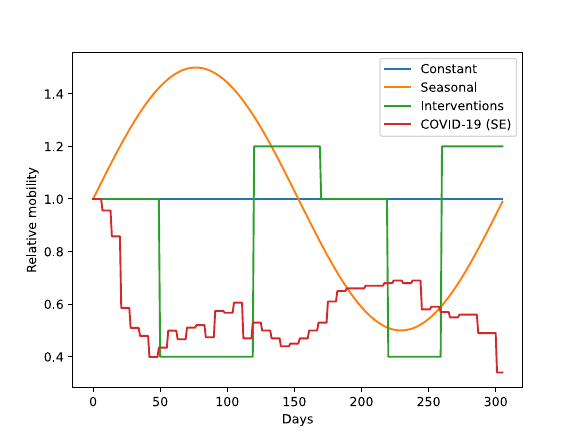}
\caption{\label{fig:mob}The different time-dependent mobilities that were used as input to CovaSim.}
\end{figure}

CovaSim allows for the generation of diverse outbreaks, thanks to the large number of parameters that can be given as input to the model, but also due to the possibility to introduce interventions that can be planned by the user. For this work we make use of non-pharmaceutical interventions in terms of a time-dependent mobility which modulates the transmission probability of all contacts. We consider four different mobilities: i) actual mobility data from Sweden during the COVID-19 pandemic, ii) seasonal mobility modelled as a sine curve, iii) rapid changes to the mobility meant to represent mandatory lock-downs and iv) a constant mobility (see fig. \ref{fig:mob}).

To begin with we used default settings for all parameters meaning that the disease being modelled is similar to COVID-19. The location was set to 'Sweden' which implies that the age distribution, contact networks and other population properties are set to match Swedish data. To reduce computational time we set the population size to $10^6$ individuals.

We assume that all severe and critical cases require hospital care and consider the daily number of critical and severe cases as the number of hospitalised each day. This time-series will be target for the forecast models.

\subsection{Parameter sensitivity analysis}
In order to generate a diverse set of outbreaks we need to define a metric that quantifies the difference between two outbreaks (in terms of time-series of hospitalised per day). Initial testing showed that it was not sufficient to simply take the absolute or squared distance between the time-series representing the hospitalisations of the two outbreaks. Given two time-series of hospitalisations $Y_1$ and $Y_2$ we opted for the following metric:
\begin{align}
    &\mathcal{L}(Y_1, Y_2) = \| \left(\| {Y_1} - {Y_2} \|_{L_1}, \| Y_1' - Y_2' \|_{L_1} , \| {Y_1''} - {Y_2''} \|_{L_1}, \frac{\max(Y_1)}{\max(Y_2)} ,\frac{\max(Y_1')}{\max(Y_2')} , \frac{\max(Y_1'')}{\max(Y_2'')} \right) \|_{L_2},
\end{align}
where $Y_i'$ and $Y_i''$ are the first and second numerical derivatives and $\| Y\|_{L_1} = \sum_i |Y^i|$ is the sum of absolute values of the vector $Y$ and $\| X \|_{L_2} = \sum_i X_i^2$. It can be noted that $\mathcal{L}$ is similar to the Sobolev norm  $\| {Y_1}  - {Y_2} \|_{W^{2, 1}} $ with the additional terms accounting for maximal difference in amplitude. 

As CovaSim has a large set of inputs parameters, a first subset of key parameters was identified: the \textit{spread} parameters and the \textit{severity} parameters, which are parameters related to disease transmission and severity.
The \textit{severity} parameters are the 4 parameters that correspond to the probability for an individual to move from one disease compartment to another. 
The \textit{spread} parameters are 9 parameters that represent the distribution of probability of the time spend by an agent in a compartment (such as infected, crictical...) once it has been entered. The value of each individual is drawn from a log-normal distribution, and the \textit{spread} parameters correspond to the mean of this log-normal distribution. All the parameters have a default value of 1, which correspond to keeping the reference value. 

This set of 13 parameters is denoted $S$. 
In order to test the forecasting models on a large, but still manageable set of different pandemics we decided to vary four parameters and that these should each take three different values in $[0.5, 1, 2]$, leading to a set of 81 pandemics. 
To select the 4 parameters in $S$ that generated the most diversity with respect to $\mathcal{L}$ out of the 13 possible is equivalent to solving the following problem: 
\begin{equation*}
s_\textrm{opt} = \underset{s \subset S , \vert s \vert =4 }{argmax  } \ \mathcal{ H}(s),    
\end{equation*}
with $ \mathcal{H}(s)= \sum_{p_1, p_2 \in \mathcal{P}_g(s)  }{\mathcal{L}_2(p_1,p_2)}$, and  $\mathcal{P}_{g}(s)$ the set of the 81 pandemics generated with the 4 parameters of $s$. However, generating an outbreak with CovaSim is time-consuming, and it is not possible to compute the diversity of each set of 4 parameters $s$ included in $S$ (the set of all 13 parameters) in reasonable time.

An MCMC algorithm \cite{boyd2009fastest} was therefore implemented, to perform a clever grid search on the different subsets $s \subset S $ of parameters. After 200 iterations of the MCMC algorithm, the set of parameters that maximised the diversity was found to be  \texttt{[sym2sev, asym2rec, rel\_symp\_prob, rel\_severe\_prob]}. 

They correspond respectively to the mean of the log-normal distribution representing the time spent in the compartment 'symptomatic' before moving into the compartment 'severe', the mean of the log-normal distribution representing the time spent in the compartment 'asymptomatic' before moving into the compartment 'recovered', to the scale factor for proportion of symptomatic cases and to the scale factor for proportion of symptomatic cases that become severe. (see \cite{kerr2021covasim}). 

Each of the 4 parameters was set to three different values: \texttt{[0.5, 1, 2]} and we considered the four time-dependent mobilities described above resulting in $81 \times 4 = 324$ different outbreaks.

\subsection{Forecast models}
In this study, we define a forecast model as a function ${h_\theta(i)}$, which equals the number of hospitalised cases at day $i$ since the outbreak began, with parameters $\theta$ that are estimated by training on the data $\mathcal{D}$.
In the training phase, $\hat{\theta}$,  an estimator of $\theta$ is computed from $\mathcal{D}$, and used for the prediction.

We considered two types of models: univariate models which are only trained on the time series we want it to predict (the number of hospitalized in our case), and multivariate models that are trained on the time series we want to predict, but also on other time series that can be relevant to predict the number of hospitalized: the mobility and incidence (new cases/day). 
%All of these models were implemented in Python, and are available on the Github repository.

Another way to classify the models is according to their underlying structure. We will consider statistical models that use time as a covariate, autoregressive models that use data from previous days as covariates and mechanistic models that are formulated as systems of coupled ordinary differential equations.

During the training or prediction phase computations sometimes fail (e.g. due to non-invertible matrices). 
Instead of outputting a missing value, we let the model output the value of the Moving average-model (see below), which can be interpreted as a naive fallback when computation fails. 

For a forecast that is made on day $i$ that reaches $r$ days into the future the model is trained on the set $\mathcal{D}_i=\{Y_j\}_{j=0}^i$. The point forecast is given by $\hat{Y}_r = {h}_{\hat{\theta}}(r)$ where $\hat{\theta}$ are the estimated parameter values of the model. To each point forecast we associate $(1-\alpha)$ confidence intervals, where $\alpha=[0.02, 0.05, 0.1, 0.2, 0.3, 0.4, 0.5, 0.6, 0.7, 0.8, 0.9]$, making the forecasts probabilistic.

\subsubsection{Statistical models}
\textbf{Exponential regression:} This models assumes that the number of hospitalised cases follows $h(t,a,b,c)=ae^{b(t-c)}$, where $\theta=(a,b,c)$ are the parameters of the model and $t$ is the number of days since the start of the outbreak. The parameters are estimated using least squares optimisation with the SciPy-function \texttt{curve\_fit} and confidence intervals are computed using the Delta method \cite{doob1935limiting}.

\textbf{Multivariate exponential regression:} This model is similar to the univariate case and assumes that $h(t,a,b,c,d,e)=ae^{(b+dm_t+ei_t)(t-c)}$ where $d$ and $e$ are additional parameters, $m_t$ is the mobility at day $t$ and $i_t$ is the incidence at day $t$. Again, we make use of least squares optimisation and the Delta method. When predictions are made we assume that the mobility and incidence take last known value for all future dates.

\subsubsection{Autoregressive models}
\textbf{Moving average:} This is the simplest of all models and serves as our baseline-model. The prediction is constant and equals the mean number of hospitalised cases in the past seven days. The confidence interval of the prediction is calculated as the standard deviation during the past seven days.

\textbf{ARIMA:} The $\textrm{ARIMA}(p, d, q)$ model is the sum of an $\textrm{AR}(p)$ and a $\textrm{MA}(q)$ model applied on the time series differentiated $d$ times. 
It follows the equation: 
\\
$Y_{t}^{d}=\alpha+\sum_{i=1}^{p}\beta_{t-i}\,Y_{t-i}^{d}\,+\,\sum_{j=1}^{q}\phi_{t-j}\,\epsilon_{t-j} \label{eq:arima},$\\

where $Y_{t}^{d}$ is the time series at time $t$, $d$ is the order of the differentiation, $\alpha$ is a constant, $p$ is the order of the autoregressive part, $q$ is the order of the moving average part and $\epsilon_{t-j}$ is the difference between the prediction of the model and the real value at time $t-j$.\\
The coefficient are estimated through maximum likelihood estimation. 
This method is implemented in the \texttt{statsmodels} library, which directly provides prediction and confidence intervals.
We performed a grid search on a single pandemic to identify the combination of parameters that would optimize the prediction accuracy. 
We found an optimal value for $p= 3, d=0, q=3$.

%This model is implemented using the \texttt{statsmodel}-package in Python. Order selection was carried out using grid search on a single outbreak, and the optimal parameters where found to be $(p,d,q)=(3,0,3)$ corresponding to an autoregressive part which uses data from the past 3 days, no differentiation, and a moving average over the last 3 days. Parameter estimation was carried out using maximum likelihood and confidence intervals were computed using the \texttt{statsmodel}-package. When predictions are made with autoregressive models (excluding the moving average model) beyond one day the predicted time-series is itself used for further predictions. 

\textbf{VAR:} 
The VAR model is a multivariate AR-model, where several variables are predicted. 
This model exploits the correlation between variables. 
Let $Y_{1,t}, ..., Y_{k,t}$ be the times series (in our case, $k=3$ and they correspond to the number of hospitalized, the number of infected and the mobility at day $t$).

\begin{equation*}
\begin{aligned}
\textrm{VAR}(p): 
\begin{pmatrix}
Y_{1,t} \\
Y_{2,t} \\
\vdots \\
Y_{k,t}
\end{pmatrix}
&=
\begin{pmatrix}
c_1 \\
c_2 \\
\vdots \\
c_k
\end{pmatrix}
+
\begin{pmatrix}
\phi_{11,1} & \phi_{12,1} & \cdots & \phi_{1k,1} \\
\phi_{21,1} & \phi_{22,1} & \cdots & \phi_{2k,1} \\
\vdots & \vdots & \ddots & \vdots \\
\phi_{k1,1} & \phi_{k2,1} & \cdots & \phi_{kk,1}
\end{pmatrix}
\begin{pmatrix}
Y_{1,t-1} \\
Y_{2,t-1} \\
\vdots \\
Y_{k,t-1}
\end{pmatrix} \\
&\quad + \cdots 
+ 
\begin{pmatrix}
\phi_{11,p} & \phi_{12,p} & \cdots & \phi_{1k,p} \\
\phi_{21,p} & \phi_{22,p} & \cdots & \phi_{2k,p} \\
\vdots & \vdots & \ddots & \vdots \\
\phi_{k1,p} & \phi_{k2,p} & \cdots & \phi_{kk,p}
\end{pmatrix}
\begin{pmatrix}
Y_{1,t-p} \\
Y_{2,t-p} \\
\vdots \\
Y_{k,t-p}
\end{pmatrix}
+
\begin{pmatrix}
\epsilon_{1,t} \\
\epsilon_{2,t} \\
\vdots \\
\epsilon_{k,t}
\end{pmatrix}
\end{aligned}
\end{equation*}
    
Again, the $\phi_{i,j,k}$ and  $c_i$ are estimated through maximum likelihood estimation with the \texttt{statsmodel} library. 
The confidence intervals are also directy provided by the library.

%This is Vector Autoregressive-model makes use of incidence and mobility data to predict future hospitalised cases. Order selection is carried out using default settings and no constant term or trend is assumed. The parameters of the model are estimated using maximum likelihood and confidence intervals are computed using the \texttt{statsmodel}-package. 

\textbf{Linear regression:} This is an AR-model where the parameters are estimated using linear regression. In order to implement this model we converted the time-series $Y_{t,  t \in \{1, \hdots n\}}$ into a training set $(X_i, Y_i)$ such that:
$\forall i \in \{1, ..., n\}, X_i = (Y_{i-1}, Y_{i-2}, ..., Y_{i-20})$, which implies that the order of the model equals $p=20$. The model was implemented using the \texttt{scikit-learn} package. 

The confidence interval for the linear regression prediction was computed as follows: Let us suppose that the data follows a linear regression model: $Y = X\beta + \epsilon$, with $Y \in \mathbb{R}^n$, $X \in \mathbb{R}^{n \times d}$, $\beta \in \mathbb{R}^d$ and $\epsilon \sim \mathcal{N}(0, \sigma^2)$.
The least square estimator of $\beta$ is $\hat{\beta} = (X^T X)^{-1} X^T Y$.
If we create a new matrix $\tilde{X}$ with unseen data, we can perform the prediction as follow: 
%If we have new data $\tilde{X} \in \mathbb{R}^{1 \times d}$ that we want to predict, the prediction is given by:

$
\begin{aligned}
    \tilde{Y} & = \tilde{X} \hat{\beta} \\
    &  = \tilde{X} (X^T X)^{-1} X^T Y \\
    &  = \tilde{X} (X^T X)^{-1} X^T (X\beta + \epsilon) \\
    &  = \tilde{X} \beta + \tilde{X} (X^T X)^{-1} X^T \epsilon.\\
\end{aligned}
$
\newline
This implies that $\tilde{Y}$ follows a normal distribution of expected value $\tilde{X} \beta$ and variance $\tilde{X} (X^T X)^{-1} \tilde{X}^T \sigma ^2$.

\textbf{Bayesian regression:} This model is similar to linear regression, and also implemented using the \texttt{scikit-learn} package, but the parameters are estimated using Bayesian ridge regression.

The Bayesian ridge regression model treats the problem of predicting the future number of severe cases as a linear regression problem, where past observations serve as covariates. The default setting is to use data stretching 20 days into the past to predict the number of severe cases tomorrow, i.e.
\[
\hat{Y}_{t+1}= w_0 + w_1 Y_{t} + ... + w_p Y_{t-19} +  \varepsilon_t
\]
where $\hat{Y}_{t+1}$ is the predicted number of severe cases at day $t$, $Y_{t-i}$ is the number of severe cases $i$ days prior, $w_i$ are the weights, which are estimated using historical data, and $\varepsilon_t$ is Gaussian noise with mean zero and variance $\lambda^{-1}$.

The weights follow a Gaussian prior:
    \[
    P(w | \alpha) \sim \mathcal{N}(0, \alpha^{-1} I),
    \]
where \(\alpha\) is the prior precision, controlling the regularization strength.
    
The likelihood of the observed data given the weights is Gaussian:
    \[
    P(y | X, w, \lambda) \sim \mathcal{N}(X \cdot w, \lambda^{-1} I),
    \].
Both the weights and the values of $\alpha$ and $\lambda$ are estimated using Bayesian inference.

\subsubsection{Mechanistic models}
\textbf{SIRH:} The SIRH-model is an extension of the classic compartmental SIR (Susceptible-Infectious-Recovered) model used to describe the spread of infectious diseases. In the SIRH model, a fourth compartment, $H$ for Hospitalised, is added. 
The evolution of the number of individuals in each compartment is described by a system of coupled ordinary differential equations: 
\begin{equation}
    \label{eq:sirh}
    \left\{
    \begin{aligned}
        &\frac{dS}{dt} = - \beta \frac{SI}{N} \\
        &\frac{dI}{dt} = \beta \frac{SI}{N} - \gamma_i I - h I \\
        &\frac{dR}{dt} = \gamma_i I + \gamma_h h \\
        &\frac{dH}{dt} = h I - \gamma_h H
    \end{aligned}
    \right.
\end{equation}
At $t=0$, the values of $(S_0, I_0, R_0, H_0)$ are fixed to $(10^6 -1, 1, 0, 0,)$, corresponding to a single infectious individual in a otherwise susceptible population. The equations were solved using the SciPy-function \texttt{odeint}. 

To train this model, we minimize the least squares error between the number of hospitalised cases in the training data and $H(t)$ with respect to $\theta = (\beta, \gamma_i, \gamma_h, h)$, using \texttt{curve\_fit}. We implemented four version of the SIRH-model in which either $\gamma_i$, $\gamma_h$ were fixed or subject to optimisation. (see the Table.\ref{tab:sirh_models_parameters}). If fixed they were set to $\gamma_i = 1/8$ and $\gamma_h=1/18$, which is in line with the mean number of days for recovery for regular and severe cases in CovaSim.

%const_gamma_i=1/8
%const_gamma_h=1/18

\begin{table}[ht]
\centering
\begin{tabular}{|c|c|c|}
\hline
 &$\gamma_i$ & $\gamma_h$ \\
\hline
SIRH1 & 1/8 & 1/18 \\
\hline
SIRH2 & 1/8 & free \\
\hline
SIRH3 & free & 1/18 \\
\hline
SIRH4 & free & free \\
\hline
\end{tabular}
\caption{Difference between the SIRH models with respect to the $\gamma$-parameters}
\label{tab:sirh_models_parameters}
\end{table}

For a prediction made at day $t$ with reach $r$ we solve the model numerically with the above mentioned initial condition with the parameters $\hat{\theta}$ computed during the training phase. The prediction is shifted so that it equals the observed number of hospitalised cases at day $t$, i.e. $H(t)=Y_t$.
The confidence interval of the prediction is computed using the Delta method and numerical differentiation. 

\textbf{SIRH-Multi:} We also implemented two SIRH-models where the mobility data influences the contact rate. We assume that $\beta$ varies with the time as a linear combination of the mobility: $\beta_t = a + b \times m_t$, similar to what was used in \cite{gerlee2021predicting}. 

We consider two versions of the SIRH-Multi model in which the values of $\gamma_h$ and $\gamma_i$ are either fixed or subject to optimisation. SIRH-Multi1 refers to the model in which $\gamma_h$ and $\gamma_i$ are free and SIRH-Multi2 refers to the model in which they are fixed to $\gamma_i = 1/8$ and $\gamma_h=1/18$.

\textbf{SEIR-Mob:} This model is adapted directly from \cite{gerlee2021predicting} and consists of an SEIR-model with time-dependent mobility of the form $\beta_t = a + b \times m_t$. The model assumes that a fraction $p$ of the infected cases become hospitalised with a delay of 21 days. To train the model, we minimize the least squares error between the number of hospitalised cases in the training data and the model solution with respect to $\theta = (a, b, p)$, using \texttt{curve\_fit}. Model prediction and confidence intervals are computed as for the SIRH-models.

\subsection{Performance metrics}
Two metrics were used to assess the performance of the models. 
The first metric is the Weighted Interval Score (WIS), which is a metric commonly used in forecast evaluation \cite{cramer2022evaluation}. 

Let $\alpha$ be in $]0, 1[$. Let $\hat{y}$ be the prediction of the model and $y$ the real value.
Let $[l, u]$ be the $(1-\alpha)$ confidence interval of the prediction.
The Interval Score ($IS$) is defined as
\begin{equation*}
IS_\alpha([l, u],  y) = \frac{2}{\alpha} \times (\mathbbm{1}_{\{y<l\}} (l-y) + \mathbbm{1}_{\{y>u\}} (y-u) + (u-l)).    
\end{equation*}
This metric consists of three terms: a term of overprediction that punishes a model with a confidence interval at level $\alpha$ which is above the real value, a term of underprediction that punishes a model whose confidence interval is under the real value, and a term of range, that punishes too wide confidence intervals. \\
Let $(\alpha_k)_{k \in \{1, \dots , K\}} \in ] 0 , 1 [ ^K $ be a sequence of significance levels. The WIS is now defined as 

\begin{equation*}
\textrm{WIS}([l, u], \hat{y},y) = w_0 |y-\hat{y}| + \sum_{k=1}^{K} w_k IS_{\alpha_k}([l, u], y),    
\end{equation*}
with weights $(w_k)_{k \in \{1, ... , K\}} \in \mathbb{R}_+ ^K $ chosen by the user.

According to previous literature \cite{cramer2022evaluation}, we decided to set \\
$(\alpha_k)_{k \in \{1, ... , K\}} = [1,0.02, 0.05, 0.1, 0.2, 0.3, 0.4, 0.5, 0.6, 0.7, 0.8, 0.9]$ and  $ \forall k \in  \{1, ... , K\}, w_k = \frac{\alpha_k}{2}$. 

The second metric is the Root Mean Square Error (RMSE). 
With the same notations as above, we define the RMSE as 
\begin{equation*}
\textrm{RMSE}(\hat{y}, y) = \sqrt{(y-\hat{y})^2}    
\end{equation*}
This metric focuses on the point prediction, and does not take into account the confidence intervals.

The models were tested on all the 324 pandemics, on 14 data points different (at days 20, 40, 60, ..., 280). 
For each individual point, the models were trained on the previous days of the pandemic. 
A 7 and 14 days ahead point prediction was computed, and $[0.02, 0.05, 0.1, 0.2, 0.3, 0.4, 0.5, 0.6, 0.7, 0.8, 0.9]$ confidence-intervals were constructed. 
The WIS and the RMSE of these predictions were then calculated. 
For the analysis of the results, we decided to remove the points for which the number of hospitalized was below 10, (i.e less than $10^{-4}$ hospitalized per 100 000 citizens).
The reason being that it is of little use to assess the performances of the models during the period in which hospitalisations and incidence are low. Not removing those points would lead to biased results, as points of low hospitalisations represent 44\% of the dataset, and we would have concluded that the best model is the one that performs well when there is no or very little transmission. 

\subsection{Classifying stages of the epidemic}
In order to compare the performance of the models at different stages of an outbreak, we classified each day of the outbreak into five categories based on the effective reproduction number $R_{\textrm{eff}}$. This number was extracted from the CovaSim simulations. 
The classification was made according Table \ref{tab:classification}. 

\begin{table}[ht]
\centering
\begin{tabular}{|c|c|}
\hline
\textbf{Condition} & \textbf{Classification} \\
\hline
$R_{\textrm{eff}} < 0.5$ & minimal transmission \\
\hline
$0.5 \leq R_{\textrm{eff}} < 0.8$ & low transmission \\
\hline
$0.8 \leq R_{\textrm{eff}} < 1.2$ & stable \\
\hline
$1.2 \leq R_{\textrm{eff}} < 3$ & high transmission \\
\hline
$R_{\textrm{eff}} \geq 3$ & very high transmission\\
\hline
\end{tabular}
\caption{Classification according to reproduction number.}
\label{tab:classification}
\end{table}

Among all 2549 points where model predictions were evaluated, 698 were classified as 'minimal transmission', 579 as 'low transmission', 388 as 'stable', 752 as 'high transmission',  132 as 'very high transmission'.

\subsection{The ensemble model}
This section describes how the ensemble forecast were computed. 
The dataset of hospitalisation curves was randomly split into a training set and evaluation set where each was assigned to the training set with probability 0.5. Training of the ensemble weights was only performed on the training set. 

\textbf{Average Ensemble (EnsAvg):} this ensemble takes an unweighted mean of all component models to calculate the point prediction. The positions of the upper and lower confidence intervals for each significance level were also computed as an unweighted mean.

\textbf{Median Ensemble (EnsMedian):}  this ensemble uses the median of all point predictions. The positions of the upper and lower confidence intervals for each significance level were also computed using the median.

\textbf{Regression Ensemble (EnsReg):} this ensemble takes a weighted mean of the component point predictions plus an intercept 
\begin{equation}\label{eq:ensreg}
Y^E_t = w_0 + \sum_{k=1}^{14} w_k \hat{Y}^k_t    
\end{equation}
The weights $w_k$ and intercept $w_0$ were obtained using linear regression where the actual hospitalisation on day $t$ is the target variable. The positions of the upper and lower confidence intervals for each significance level were also computed according to \eqref{eq:ensreg}.

\textbf{RMSE-optimised Ensemble (EnsRMSE):} This ensemble is similar to EnsReg, but considers weights that sum to unity and no intercept. The weights are obtained by minimising the function
\begin{equation}\label{eq:ensrmse}
E_\textrm{RMSE}(\mathbf{w}) = \sum_{i=1}^{N_T} \sum_{j=1}^{15} \textrm{RMSE}({Y}_{i,j},\sum_{k=1}^{14} w_k \hat{Y}^k_{i,j})
\end{equation}
subject to the constraint $\sum_k w_k=1$. Here the double sum runs over all outbreaks in the training set and across all forecasts within a given outbreak, $\hat{Y}_{i,j}$ is the observed number of hospitalised cases and $\hat{Y}^k_{i,j}$ is the forecast of model $k$. The opimisation problem is solved using the SciPy-function \texttt{minimize} with the trustconst-method and an initial guess equal to $1/14$ for all $w_k$.

\textbf{WIS-optimised Ensemble (EnsWIS):} This ensemble is similar to EnsRMSE with the difference that it is the WIS which is optimised. Therefore we minimise the function
\begin{equation}\label{eq:enswis}
E_\textrm{WIS}(\mathbf{w}) = \sum_{i=1}^{N_T} \sum_{j=1}^{15} \textrm{WIS}([\sum_{k=1}^{14} w_k l^k_{i,j}, \sum_{k=1}^{14} w_k u^k_{i,j}],\hat{Y}_{i,j},\sum_{k=1}^{14} w_k {Y}^k_{i,j}))
\end{equation}
subject to the constraint $\sum_k w_k=1$. The opimisation problem is solved using the SciPy-function \texttt{minimize} with the trustconst-method and an initial guess equal to $1/14$ for all $w_k$.

\textbf{Rank-based Ensemble (EnsRank):} this ensemble changes its weights depending on the current effective reproductive number. For each of the five regimes defined in fig.\ \ref{tab:classification} we pick the five models with the smallest average rank $r_1<r_2<..<r_5$. We now set $\tilde{w}_k=1/r_k$ and normalise the weights so that the sum to unity according to 
\begin{equation*}
w_k=\tilde{w}_k/\sum_{i=1}^5\tilde{w}_i.
\end{equation*}
The forecast of this ensemble equals
\begin{equation*}
Y^R_t = \sum_{k=1}^5 w_k Y^k_t
\end{equation*}
where the sum runs over the top-ranked models given the current effective reproduction number.

\subsection{Ensemble variance}
To investigate the relationship between the variation of component model forecasts and ensemble error we calculated the coefficient of variation for all ensemble forecast across all outbreaks. The coefficient of variation is defined according to $CV=\sigma/\mu$ where $\mu=\frac{1}{14} \sum_{k=1}^{14}\hat{Y}^k$ is the mean prediction and $\sigma=\sqrt{\frac{1}{14} \sum_{k=1}^{14}(\mu- \hat{Y}^k)^2}$. The ensemble error was calculated for the median ensemble using the Mean Absolute Percentage Error $\textrm{MAPE}_t=|Y_t-Y^M_t|/Y_t$, where $Y_t$ is the outcome and $Y^M_t$ is the median forecast at time $t$. For the analysis we exclude points where $Y_t < 10$.

\subsection{Evaluation on Swedish data}
Data on the number of hospitalised COVID-19 patients in Sweden during 2020 was obtained from Our World in Data (https://ourworldindata.org/coronavirus). In order to obtain good estimates of the effective reproduction number for COVID-19 during the pandemic in Sweden in 2020 we estimated incidence based on COVID-19 deaths according to the method presented in \cite{wacker2021estimating}. Data on deaths were obtained from the Swedish National Board of Welfare. The effective reproduction number was estimated using the method described in \cite{cori2013new}, with which estimation of $R_\textrm{eff}$ on a given day only uses historical data.
The generation time distribution required for the estimation was obtained from \cite{knight2020estimating}. 

\section*{Author Contributions}
Grégoire Béchade: Methodology, Software, Formal analysis, Investigation, Writing - Original Draft, Writing - Review \& Editing. Tobjörn Lundh: Methodology, Writing - Review \& Editing. Philip Gerlee: Conceptualization, Methodology, Software, Formal analysis, Investigation, Writing - Review \& Editing, Visualization, Supervision.

%--- Section ---%

%--- Section ---%
\section*{Data Availability}
All code and data used in this study is available at \href{https://github.com/philipgerlee/Evaluation-of-respiratory-disease-hospitalisation-forecasts-using-synthetic-outbreak-data}{GitHub}.

%--- Section ---%
\section*{Acknowledgments}
The authors would like to thank Matthew Biggerstaff at the Centers for Disease Control and Prevention for suggesting the link between ensemble variance and accuracy.

%-------------------------------------------
% References
%-------------------------------------------

% Print bibliography
%\printbibliography
%\bibliography{references}
%\bibliographystyle{plain}

\end{document}

\maketitle

\begin{figure}[H]
\centering
\includegraphics[width=1\linewidth]{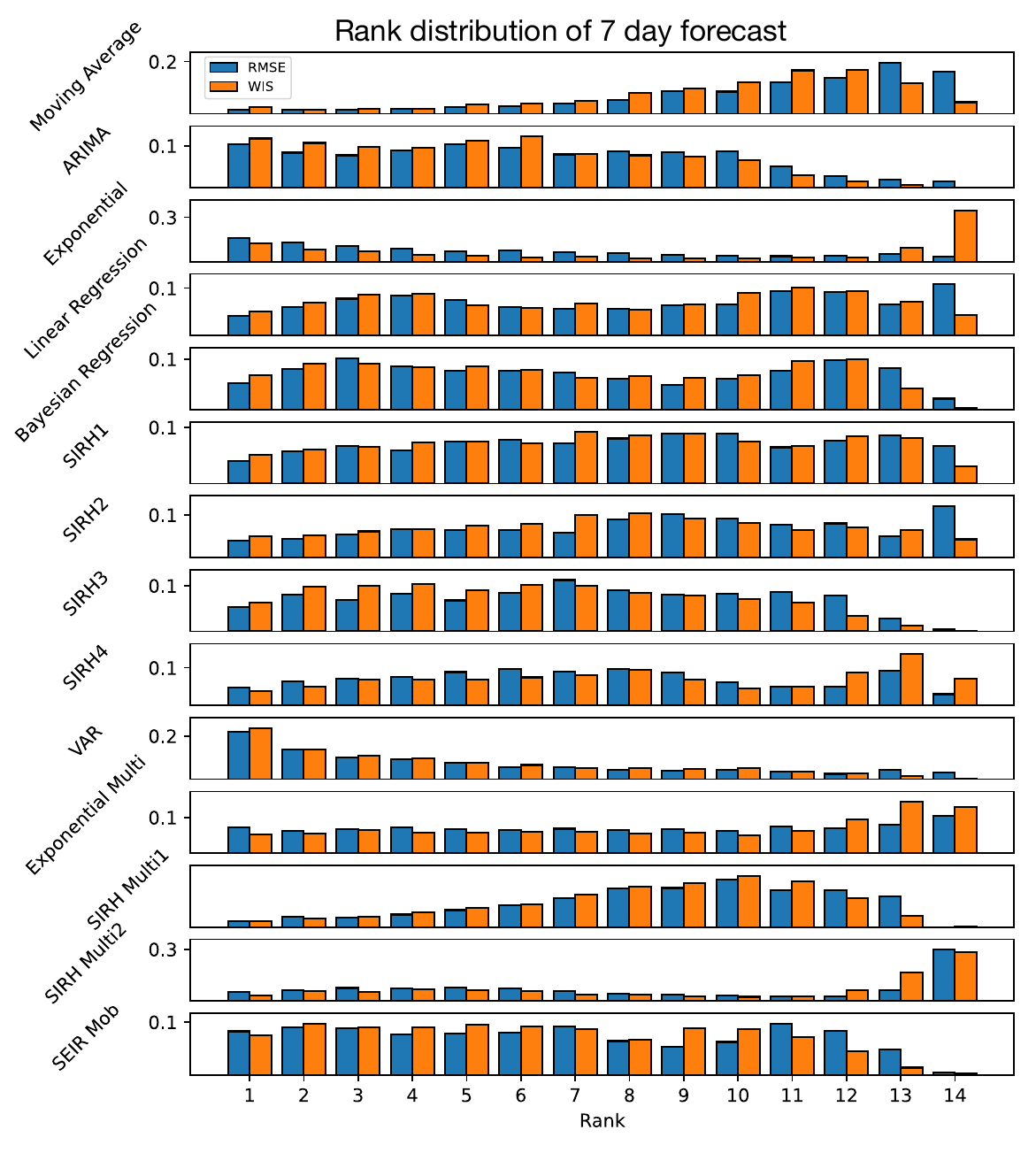}
\caption{\label{fig:overview}Distribution of rankings of the models for all points for 7-day forecasts with respect to both RMSE and WIS.}
\end{figure}
\clearpage

\begin{figure}[H]
\centering
\includegraphics[width=1\linewidth]{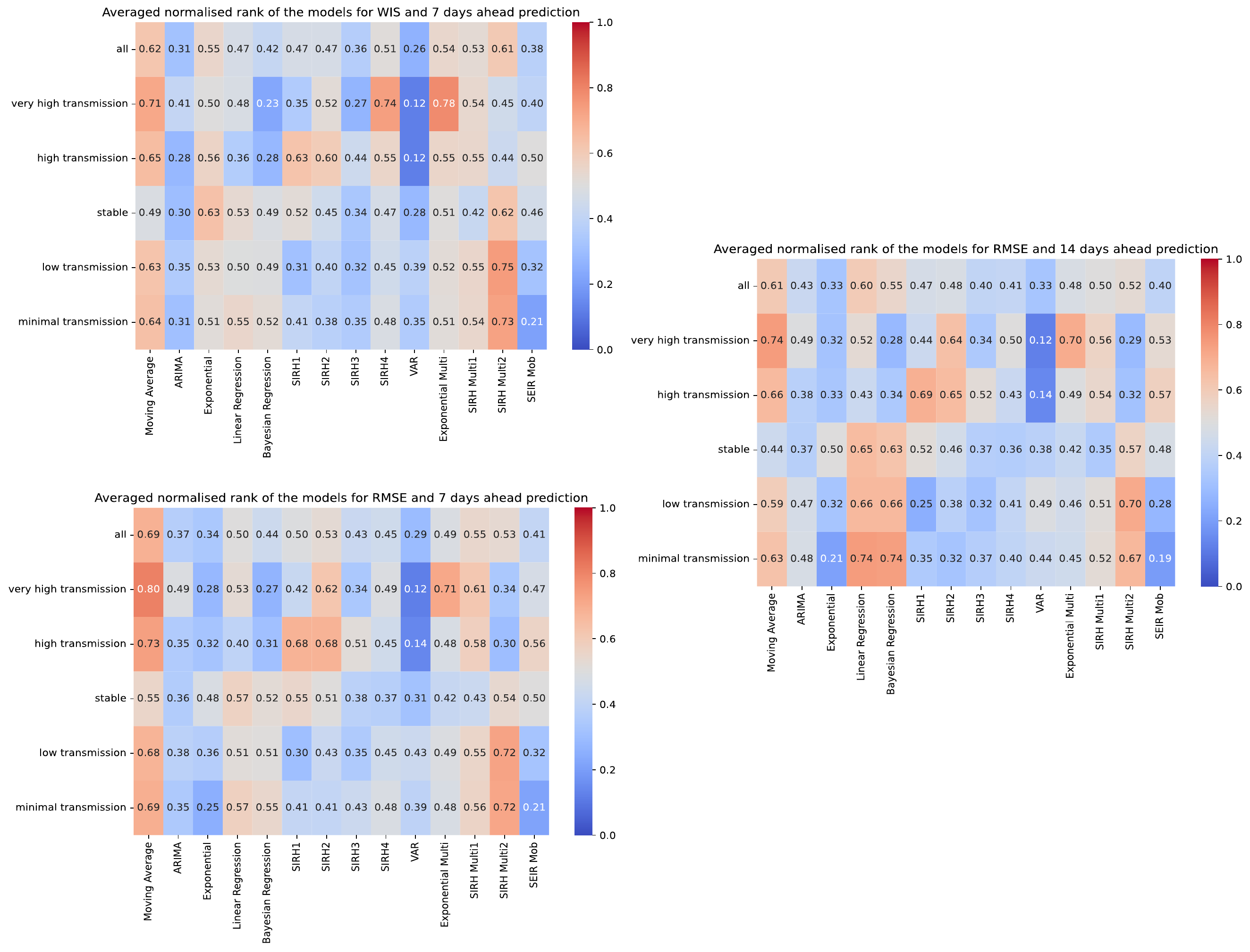}
\caption{\label{fig:overview}Heatmap of model performance based on RMSE and WIS for 7- and 14-day forecasts. The day of forecast is classified according to the current effective reproduction number according to: minimal transmission ($R_{\textrm{eff}} < 0.5$), low transmission ($0.5\leq R_{\textrm{eff}} < 0.8$), stable ($0.8\leq R_{\textrm{eff}} < 1.2$), high transmission ($1.2\leq R_{\textrm{eff}} < 3$) and very high transmission $R_{\textrm{eff}}\geq 3$.}
\end{figure}
\clearpage

\begin{figure}[H]
\centering
\includegraphics[width=1\linewidth]{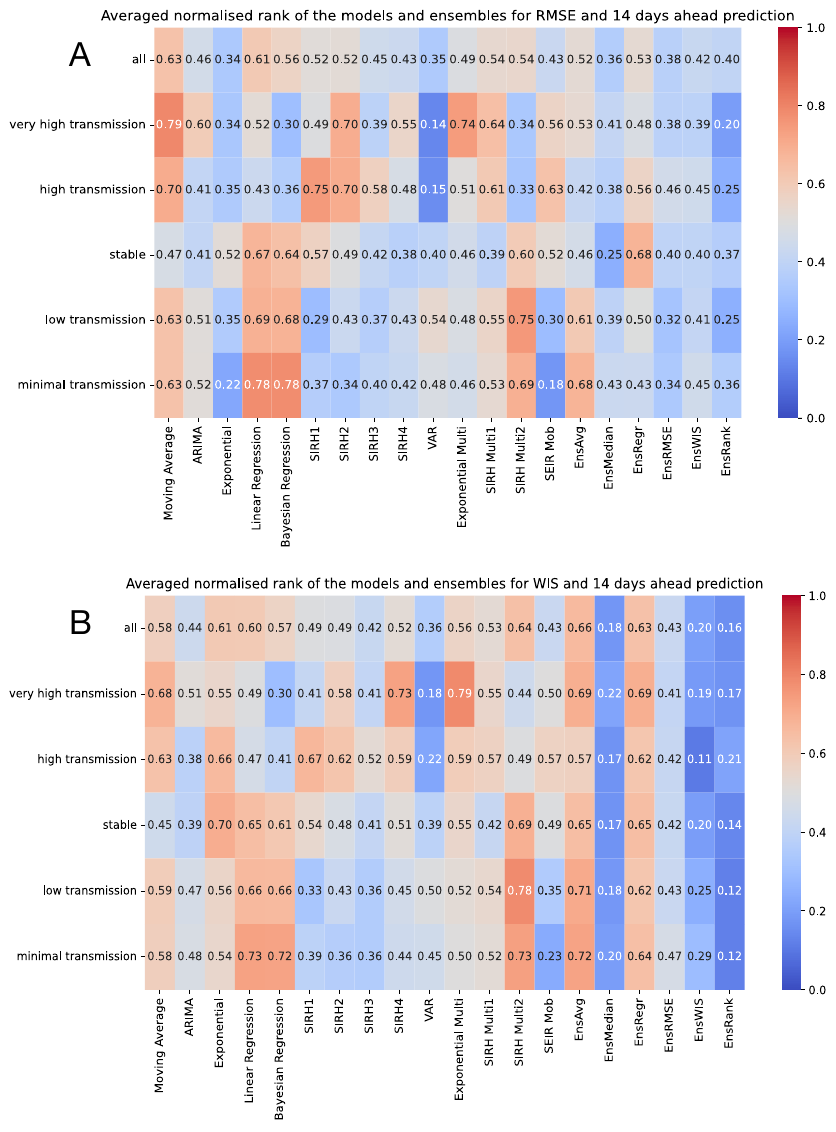}
\caption{\label{fig:overview}The average ensemble rank as compared to both component models and the six ensembles for A) RMSE and B) WIS.}
\end{figure}
\clearpage

\begin{figure}[H]
\centering
\includegraphics[width=1\linewidth]{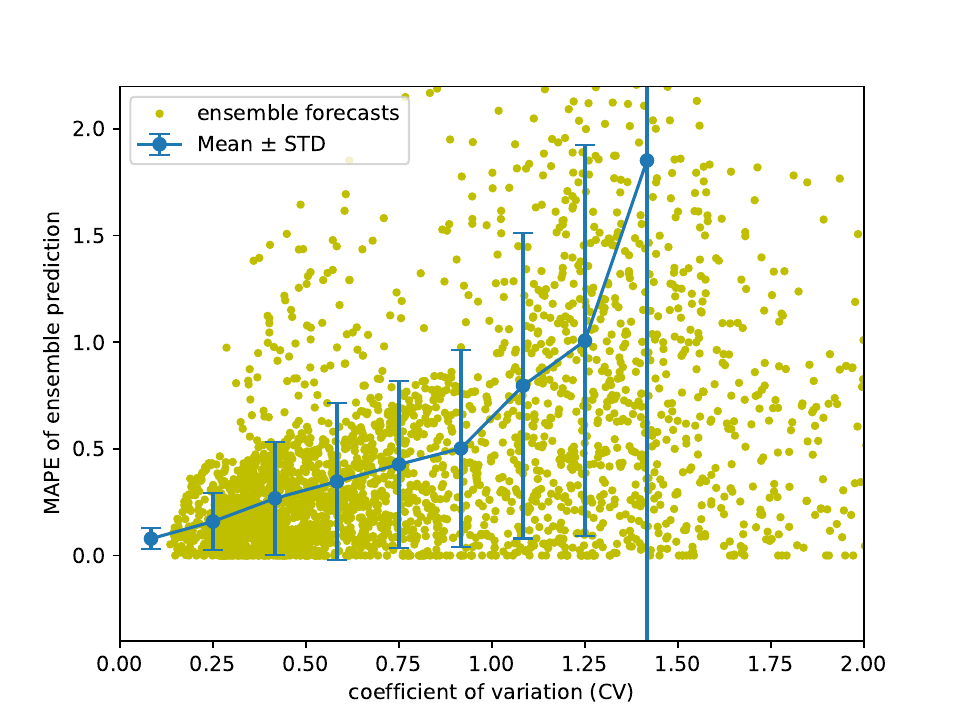}
\caption{\label{fig:overview}The coefficient of variation (CV) of the individual model predictions for 14-day forecasts and the corresponding error (MAPE) of the Median Ensemble prediction. The solid line shows the mean MAPE in each bin and the error bars correspond to one standard deviation.}
\end{figure}
\clearpage